\documentclass[a4paper,11pt]{article}
\usepackage{pos}

\title{Jet medium modifications}

\author*[a]{Carlota Andres}

\affiliation[a]{CPHT, CNRS, Ecole polytechnique,\\
 Institut Polytechnique de Paris, 91120 Palaiseau, France}

\emailAdd{carlota.andres-casas@polytechnique.edu}

\abstract{Since the start of the heavy-ion collision programs at the Relativistic Heavy Ion Collider and the Large Hadron Collider, the study of jet modifications resulting from their interactions with the produced QCD matter has provided a unique tool to investigate and characterize the properties of the quark-gluon plasma. In this mini-overview, I will present the recent theoretical advancements in describing and understanding the modifications of jets within a QCD medium.}

\FullConference{HardProbes2023\\
 26-31 March 2023\\
 Aschaffenburg, Germany\\}

\begin{document}
\maketitle

\section{Introduction}

In high-energy nuclear collisions, the occasional occurrence of a hard scattering between the incoming partons leads to the production of highly virtual partons. These high-energy partons then traverse all the stages of the system's evolution, including the quark-gluon plasma (QGP) phase, eventually fragmenting into collimated sprays of hadrons known as jets. The interactions between heavy-ion jets and the QCD matter generated in these collisions give rise to distinct modifications compared to proton-proton jets. This phenomenon, widely known as jet quenching, has become a crucial avenue for studying the QGP. 

\section{Energy loss}

Jet quenching was first established at the Relativistic Heavy Ion Collider (RHIC) in the context of single-inclusive hadron spectra and high-$p_T$ hadron correlations, with the observed suppression indicating the energy loss experienced by the leading particle as it traverses the QGP \cite{STAR:2003pjh}. In typical events, the leading parton is either a light quark or a gluon, and its energy loss is primarily driven by its frequent soft interactions with the medium which generate collinear (medium-induced) radiation. Consequently, the early emphasis within the jet quenching community was on computing the spectrum of medium-induced radiation and its applications to energy loss observables. 

A complete description of a medium-induced gluon emission off a highly energetic parton requires the resummation of multiple scatterings, given the important role played by interference effects among scatterings resulting into the well known Landau-Pomeranchuk-Migdal (LPM) effect. This resummation can be formally performed within the BDMPS-Z framework \cite{Baier:1996kr,Zakharov:1996fv}. To successfully incorporate all multiple scatterings into the well-known BDMPS-Z compact formula, several assumptions are made. Firstly, it is assumed that the opening angle of the radiation is small, and the emission vertices are described by leading-order DGLAP splitting functions. Secondly, the interactions between the parton and the medium are considered instantaneous and mediated by soft gluons, which only transfer transverse momenta to the probe. In addition, a common simplifying assumption is to treat the radiated gluon as soft, allowing for its interpretation as energy loss of the emitter. This assumption is typically expressed as $\omega \equiv zE \ll E$, where $\omega$ and $E$ represent the energies of the radiated gluon and the initial parton, respectively. By considering this limit, the calculation is simplified, and the dominant contribution to the spectrum is obtained due to the soft divergence of the DGLAP splitting functions for gluon radiation. It is important to note that while implementing this soft limit is not strictly required for the derivation of the BDMPS-Z spectrum, its relaxation while still retaining information on the emission angle $\theta$ has proven to be challenging \cite{Blaizot:2012fh, Apolinario:2014csa}. 

\begin{figure}
\vspace{-5mm}
\centering
\includegraphics[scale=0.42]{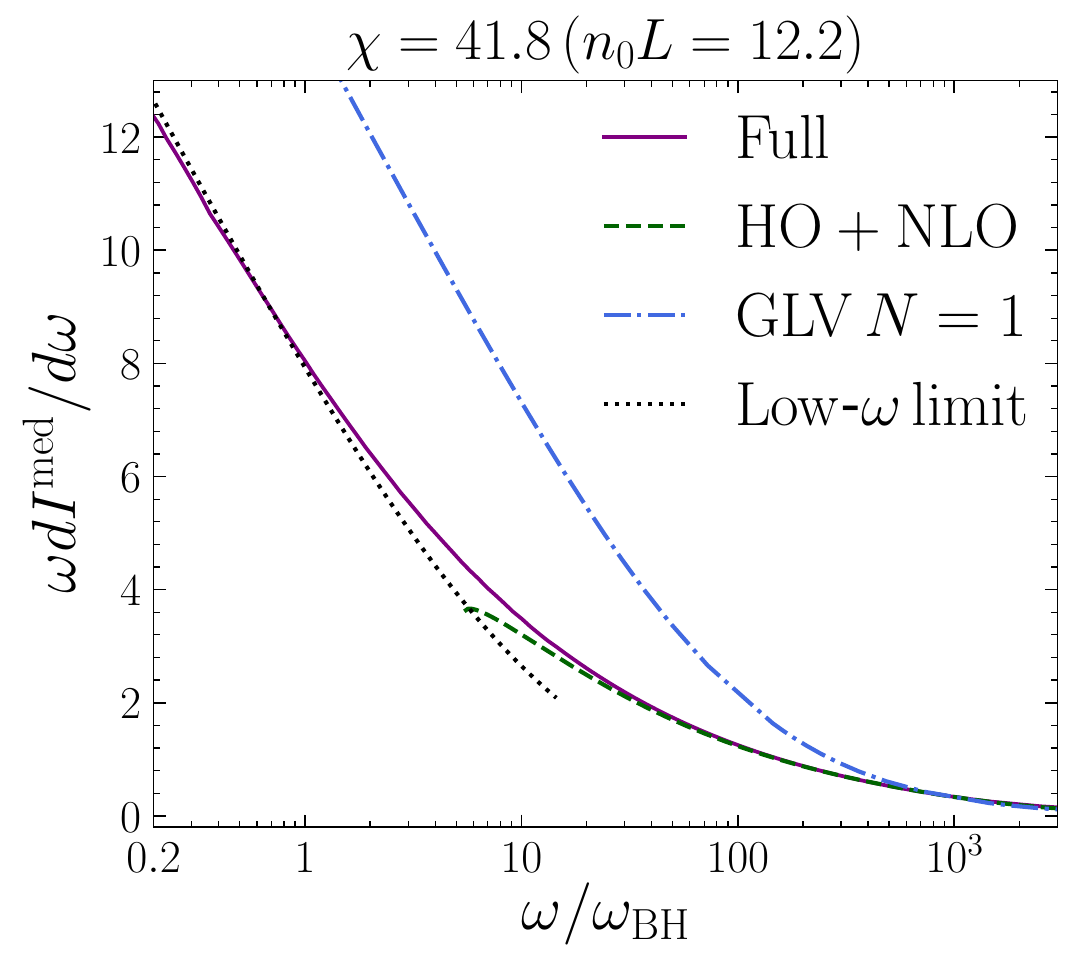}
\caption{Full medium-induced gluon spectrum from \cite{Andres:2020vxs} (magenta solid line) compared to the harmonic oscillator+NLO approximation from \cite{Mehtar-Tani:2019tvy,Barata:2020sav} (green dashed), first opacity result (blue dash-dotted) \cite{Gyulassy:1999zd}, and the low energy limit of the full resummation described in section~4 of \cite{Andres:2020kfg} (black dotted) as a function of  $\omega$ rescaled by the energy scale $\omega_{\mathrm{BH}}$. Figure taken from \cite{Andres:2020kfg}, to which we refer the reader for details.} 
\label{fig:spectra}
\end{figure}

Historically, the evaluation of the BDMPS-Z formula has been limited to a few specific scenarios. One approach involves truncating the resummation series at first order in an opacity expansion \cite{Gyulassy:1999zd}, while another method resumes all multiple scatterings but under the assumption that the transverse momentum transfers follow a Gaussian profile (known as the harmonic oscillator approximation). The determination of the spectrum also simplifies for an infinite static medium \cite{Arnold:2002ja}. In recent years, significant efforts have been dedicated to systematically improving our understanding of medium-induced emissions. Both analytical and numerical advancements have enabled the evaluation of the BDMPS-Z spectrum beyond these simplified scenarios \cite{Caron-Huot:2010qjx, Feal:2018sml,Mehtar-Tani:2019tvy,Barata:2020sav,Andres:2020vxs,Andres:2023jao,Stojku:2023ell}, allowing us to quantify their kinematic ranges of validity for an idealized static QGP \cite{Andres:2020kfg}. This progress is illustrated in figure~\ref{fig:spectra}, where the first opacity result approximates the high energy tail of the spectrum well, but tends to overestimate it at lower gluon energies due to the absence of interferences among scatterings.  These evaluations were agnostic to the specifics of the collision rate or parton-medium interaction model entering the spectrum expression, provided it exhibits the expected Coulomb-like screening behavior for large transverse momentum transfers. Notably, recent studies \cite{Schlichting:2021idr} have revealed that incorporating a collision rate with proper matching to the infrared sector \cite{Moore:2019lgw} has a large impact on the spectrum. 

\begin{figure}
\vspace{-5mm}
\centering
\includegraphics[scale=0.55]{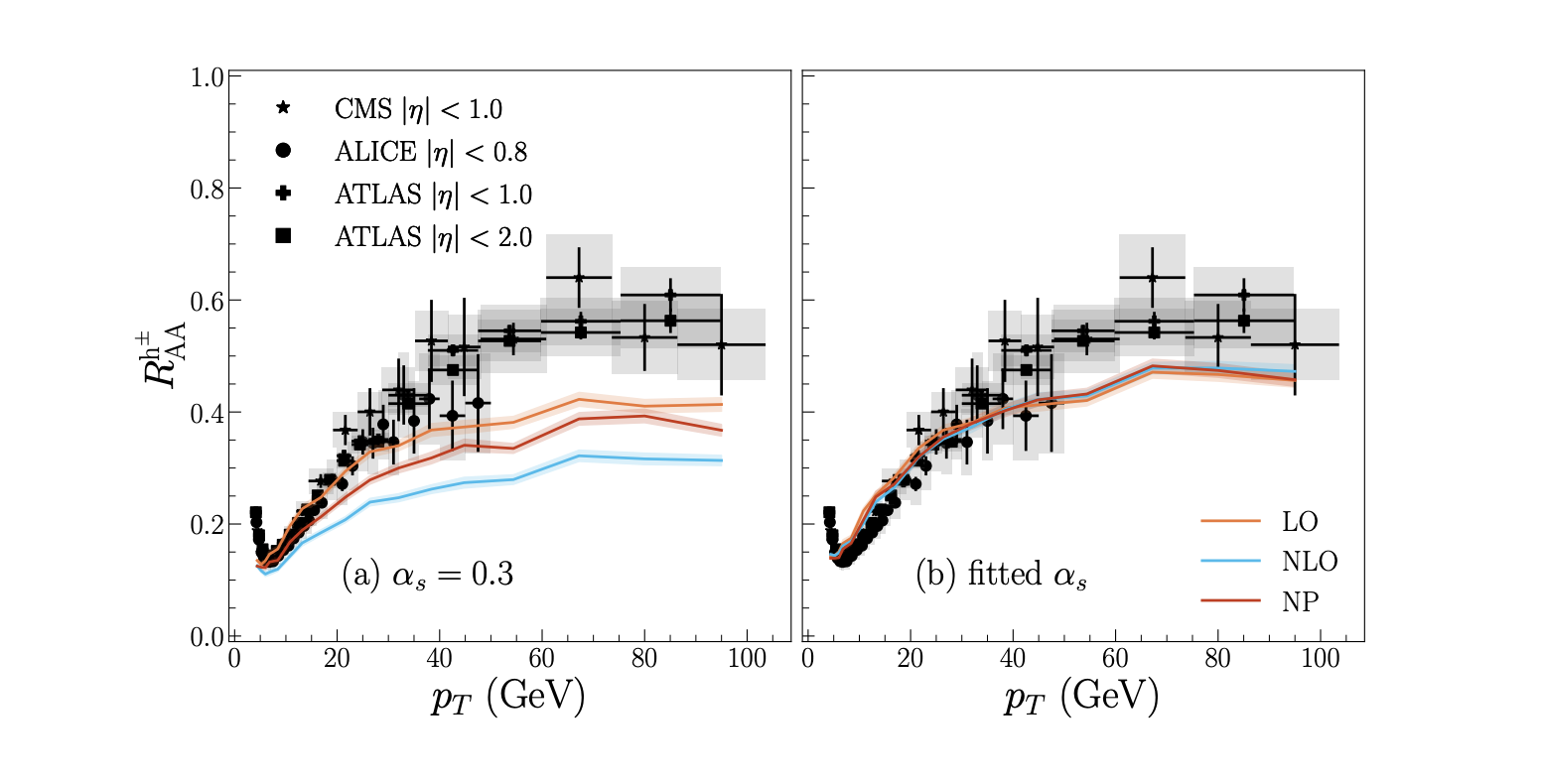}
\caption{Charged hadron $R_{\rm AA}$ as a function of $p_T$ computed using MARTINI \cite{Schenke:2009gb} with AMY (infinite length) rates \cite{Arnold:2002ja} using either a LO (orange) collision rate, a NLO collision rate (blue), or a collision rate matched to the infrared Lattice result in \cite{Moore:2019lgw} (red). The left panel uses $\alpha_s=0.3$ for the three curves, while the right panel uses fitted values of $\alpha_s$, yielding: $\alpha_s=0.28$ (orange), $\alpha_s=0.24$ (blue), and $\alpha_s=0.26$ (red).  Figure taken from  \cite{Yazdi:2022bru}, to which we refer the reader for further details.} 
\label{fig:raa}
\end{figure}

While these advancements have significantly deepened our understanding of the in-medium emission process, their implications on jet quenching  phenomenological analyses are yet to be fully explored. In fact, for some energy loss observables, the differences among various approaches might potentially be absorbed into the fitted values of free parameters \cite{Yazdi:2022bru}, typically the jet transport coefficient $\hat q$ or the strong coupling $\alpha_s$, depending on the formalism. This is illustrated in figure~\ref{fig:raa}, where we observe that for the $0-5\%$ centrality class in $\sqrt{s_{\rm NN}}= 2.76$ TeV Pb-Pb collisions, the curves representing different collision kernel models collapse onto each other and become indistinguishable when the strong coupling value is separately fitted to the charged hadron $R_{\rm AA}$ data for each of them.  This outcome is not entirely surprising, as the $R_{\rm AA}$ dependence on $p_T$ is primarily governed by the steeply decreasing (with $p_T$) hard hadron production spectrum. Therefore, it would be highly desirable to analyze the impact of these developments on the theory of medium-induced radiation on other more differential observables.

Furthermore, recent endeavors have focused on investigating the impact of the transverse flow or transverse inhomogeneities within the QGP on medium-induced radiation and broadening. To address these aspects, it becomes necessary to generalize the original BDMPS-Z formalism, since in the high-energy (\emph{eikonal}) limit for the emitter  in which it was derived the longitudinal and transverse degrees of freedom are completely decoupled. The relaxation of the eikonal limit allows to compute corrections due to transverse flow and/or transverse density gradients which are  proportional to  inverse powers of the energy of the emitted gluon \cite{Sadofyev:2021ohn,Barata:2022krd,Andres:2022ndd,Barata:2022utc,Barata:2023qds}. These \emph{sub-eikonal} transverse corrections  are expected to cause the jets to drift in the direction of the transverse flow or temperature gradient, resulting in an asymmetry in their transverse distribution \cite{Antiporda:2021hpk,Fu:2022idl}.

\section{From energy loss to jet substructure}

Recent years have witnessed remarkable progress  in the measurement of increasingly differential jet observables in heavy-ion collisions, see e.g.~\cite{Connors:2017ptx} for a comprehensive review. Of particular interest is the study of modifications to the inner structure of heavy-ion jets, as it provides insights into the dynamics of the QGP at its different stages, which is reflected in the various scales of the substructure of the jet. However, the theoretical interpretation of jet substructure observables remains challenging  due to the interplay of multiple phenomena, including color coherence \cite{Mehtar-Tani:2010ebp} and medium response \cite{Cao:2020wlm} effects.

Medium-induced emissions, when radiated at small angles, can also modify the inner structure of jets, making the precise description of in-medium emissions crucial for fully exploiting jet substructure observables as probes of the QGP. A key aspect in this regard is to accomplish the complete calculation of medium-induced splittings, taking into account the differential dependence on both the relative angle $\theta$ and the fraction of longitudinal momentum $z$ carried by the emitted parton, beyond the soft limit in which $z\rightarrow 0$. To tackle this issue, two available approaches have been considered: performing an expansion in opacity or resumming all scatterings using a \emph{semi-hard} approximation. At first order in opacity, the double-differential spectrum beyond the soft limit, i.e for finite (non-zero) $z$, is well known \cite{Ovanesyan:2011xy}. Recursive formulas have been derived for higher-order opacity terms, enabling the numerical evaluation at second order \cite{Sievert:2018imd, Sievert:2019cwq}. However, it is important to note that truncating the in-medium splitting in opacity leads to unitarity issues, which results in negative cross sections. Alternatively, the \emph{semi-hard} approximation within the BDMPS-Z framework assumes that all partons propagate along straight-line trajectories while undergoing color rotations \cite{Dominguez:2019ges,Isaksen:2020npj}. While this simplification enables the calculation of the double-differential spectrum with resummation of all multiple scatterings for finite $z$, it neglects certain broadening effects. These two approximations of the double-differential spectra have been employed, for instance, to compute the two-point energy correlator of a heavy-ion jet \cite{Andres:2022ovj,Andres:2023xwr}, which is a novel and promising substructure observable in heavy-ion physics. Notably, the first-ever preliminary measurements of this observable in p-p jets were presented by the STAR and ALICE collaborations at this conference \cite{Tamis:HP2023, Cruz-Torres:HP2023}, showing a clear (and expected)  angular separation between its non-perturbative and perturbative regimes \cite{Komiske:2022enw}.  

Recently, a significant development has been the computation of the double-differential spectrum incorporating the resummation of multiple scatterings for finite $z$, without relying on the semi-hard approximation, for specific splittings. One of these cases is the  in-medium  $g \rightarrow c\bar c$ splitting, which was obtained in the large number of colors (large-$N_c$) limit within the harmonic oscillator approximation \cite{Attems:2022ubu}, revealing an enhancement of the number of $c\bar c$ pairs in heavy-ion jets compared to p-p jets. Notably, the computation of this particular splitting at large $N_c$ is slightly less complicated compared to other $1\rightarrow 2$ splittings, as it does not involve the calculation of 4-point correlators of Wilson lines (quadrupole). The other case is the $\gamma \rightarrow q\bar q$ splitting obtained in \cite{Isaksen:2023nlr}, which already involves the calculation of the quadrupole at large $N_c$. In this large-$N_c$ limit and under the harmonic oscillator approximation, the quadrupole can be analytically obtained. More interestingly, the authors in \cite{Isaksen:2023nlr} have developed a numerical method based on solving a system of coupled Schr\"odinger equations, which allows for the calculation of the quadrupole for finite numbers of colors. This recipe could potentially be applied to the $q \rightarrow qg$ and $g \rightarrow gg$ processes as well, providing an opportunity to evaluate the accuracy of the commonly used approximations to obtain the double-differential spectrum for these emissions, including the \emph{semi-hard} approximation. It is worth noting, however, that the computational complexity of the approach in \cite{Isaksen:2023nlr} would considerably increase for these particular splittings.

Up to this point, our discussion has focused on single in-medium gluon emissions. However, over the past years there has been an ongoing effort to determine the in-medium double-gluonic splitting ($g \rightarrow ggg$), see \cite{Arnold:2022mby,Arnold:2023qwi} and references therein. The ultimate goal of these calculations is to determine whether in-medium showers can be treated as a sequence of independent $1\rightarrow 2$ splittings or if there is a significant quantum overlap between successive splittings. These developments suggest, albeit with some caveats, that for gluonic cascades the overlapping effects among multiple splittings are expected to be small \cite{Arnold:2023qwi}. This finding implies that the assumed Markovian nature of in-medium showers is not significantly challenged.

\section{The role of the initial stages on jet quenching phenomena}

Given that jets originate from hard partons produced in the hard process occurring in the initial collision, they witness the full space-time evolution of the produced QCD matter, including the \emph{initial stages}. However, most of jet quenching phenomenological studies typically neglect any quenching effects prior to the hydrodynamization phase. Recent studies have shown that certain energy loss observables are indeed sensitive to these early stages \cite{Andres:2019eus,Stojku:2020wkh,Andres:2022bql}. Nevertheless a comprehensive understanding of the impact of these initial stages on jet quenching observables in heavy-ion collisions is still lacking. Importantly, understanding jet quenching in the early stages becomes crucial to interpret the apparent lack of energy loss in small collisions systems (p-A and high multiplicity p-p), where these initial stages constitute a significantly larger fraction of the overall system's evolution compared to A-A collisions.

In the last few years, there has been a growing effort to compute the momentum broadening of both jets and heavy quarks during the early stages of heavy-ion collisions \cite{Ipp:2020mjc,Ipp:2020nfu,Carrington:2021dvw,Carrington:2022bnv,Avramescu:2023qvv,Boguslavski:2020tqz,Boguslavski:2023alu,Du:2023izb}. These advancements encompass the glasma phase \cite{Ipp:2020mjc,Ipp:2020nfu,Carrington:2021dvw,Carrington:2022bnv,Avramescu:2023qvv}, which occurs shortly after the initial collision and is characterized by highly occupied classical gluonic fields, as well as the kinetic theory stage \cite{Boguslavski:2020tqz,Boguslavski:2023alu,Du:2023izb} where the system is described as an interacting gas composed of gluon and quark quasiparticles. These calculations revealed that the jet quenching parameter $\hat q$ in these pre-hydrodynamization stages is relatively large, comparable to or even larger than in the QGP phase, suggesting that these early stages might have a significant effect on jet quenching phenomena. Moreover, both in the glasma and kinetic phases $\hat q$  has been found to be anisotropic, with a larger magnitude along the beam direction. This anisotropy would lead to the polarization of in-medium emissions from initially unpolarized emitters. However, the polarization may be washed out by subsequent branchings making it uncertain whether a net jet's polarization may survive the QGP phase \cite{Hauksson:2023tze}.

\section{Conclusions}

In recent years, there have been significant advancements in the theoretical description of the modifications experienced by jets and high-$p_T$ hadrons in heavy-ion collisions with respect to p-p collisions. These developments have the potential of serving as valuable benchmark for current jet quenching Monte Carlo (MC) approaches. While jet quenching MC simulations are highly useful, they currently rely on various modeling assumptions and approximations that may obscure the interpretation of the underlying physical phenomena (see e.g.~\cite{Apolinario:HP2023}). The ongoing progress based on first-principles QCD calculations summarized in this overview may provide an opportunity to guide and test future MC developments.

\section{Acknowledgements}
I would like to express my gratitude to the Hard Probes 2023 Organizing Committee for the invitation to present an overview on the medium modification of jets in heavy-ion collisions. C.A. has received funding from the European Union's Horizon 2020 research and innovation program under the Marie Sklodowska-Curie grant agreement No 893021 (JQ4LHC). This work was supported by OE Portugal, Funda\c{c}\~ao para a Ci\^encia e a Tecnologia (FCT), I.P., projects EXPL/FIS-PAR/0905/2021.

\end{document}